# Altmetrics can capture research evidence:
## a study across types of studies in COVID-19 literature

Pilar Valderrama-Baca[1](0000-0002-5183-7973), Wenceslao Arroyo-Machado[1](0000-0001-9437-8757),
Daniel Torres-Salinas[1](0000-0001-8790-3314)

[1]Departamento de Información y Comunicación, Facultad de Comunicación e Información, Universidad de Granada

**Abstract:**

There has been a proliferation of descriptive for COVID-19 papers using altmetrics. The main objective of this study is to analyse whether the altmetric mentions of COVID-19 medical studies are associated with the type of study and its level of evidence. Data were collected from PubMed and Altmetric.com databases. A total of 16,672 study types (e.g., Case reports or Clinical trials) published in the year 2021 and with at least one altmetric mention were retrieved. The altmetric indicators considered were Altmetric Attention Score (AAS), News mentions, Twitter mentions, and Mendeley readers. Once the dataset had been created, the first step was to carry out a descriptive study. Then a normality hypothesis was contrasted by means of the Kolmogorov-Smirnov test, and since it was significant in all cases, the overall comparison of groups was performed using the non-parametric Kruskal-Wallis test. When this test rejected the null hypothesis, pair-by-pair comparisons were performed with the Mann-Whitney U test, and the intensity of the possible association was measured using Cramers V coefficient. The results suggest that the data do not fit a normal distribution. The Mann-Whitney U test revealed coincidences in five groups of study types, the altmetric indicator with most coincidences being news mentions and the study types with the most coincidences were the systematic reviews together with the meta-analyses, which coincided with four altmetric indicators. Likewise, between the study types and the altmetric indicators, a weak but significant association was observed through the chi-square and Cramers V. It is concluded that the positive association between altmetrics and study types in medicine could reflect the level of the pyramid of scientific evidence.

## 1. Introduction

COVID-19 has affected society worldwide, as an unprecedented challenge (**Chriscaden**, 2020). The fact that this exceptional situation has greatly impacted science is attested to by an exponential explosion of scientific literature (**Torres-Salinas**, 2020; **Torres-Salinas** *et al.*, 2020). Along with the growth of publications, there were international calls for cooperation and openness of research to find a solution. This meant a unique opportunity for an open science revolution, which eventually faded away (**Brainard**, 2021). In addition, given the global impact of the pandemic and its effects on multiple aspects of society, COVID-19 attracted the attention of researchers in areas beyond Medicine from the very beginning (**Aristovnik** *et al.*, 2020). Thus, COVID-19 has become a consolidated global research front, of great interest to the scientometric community, among others.

Bibliometrics properties of this explosion of publications have been studied in detail and particularities being highlighted; **Zhang** *et al.* (2020) studied the early global response of researchers in comparison with other outbreaks; **Nane** *et al.* (2022) developed predictive models of expected publications to show the exceptional growth patterns of the scientific literature on COVID-19; **Pinho-Gomes** *et al.* (2020) analysed the gender gap in the early literature and found that only a third of the authors were women; **Zhang** *et al.* (2021) detected certain changes in research interests after the pandemic peak, while others resume previous research lines. Furthermore, the impact of these new publications on bibliometric indicators has been studied (**Fassin**, 2021). There has also been a proliferation of descriptive studies examining new sources and datasets related to COVID-19. For instance, **Colavizza** *et al.* (2021) explored and detailed the content of new bibliographic data sources, whereas **Kousha & Thelwall** (2020) compared the coverage of the scholarly databases on COVID-19 publications, pointing to *Dimensions* as the most comprehensive.

Furthermore, different proposals have been developed to understand how these new publications are shared and discussed in different social media through "altmetrics" (**Priem**, 2014). These social media metrics have proven useful for understanding aspects of science communication beyond traditional channels (**Arroyo-Machado** *et al.*, 2021). Regarding COVID-19 related research, **Kousha and Thelwall** (2020) studied the altmetric impact of COVID-19 publications on different social media and found that early altmetric mentions such as tweets reflect a positive relationship with later citations. *Twitter* is precisely one of the main social media studied, the object of numerous studies exploring diverse communities of users and interactions produced around anti-vaccine movements and disinformation (**Hayawi** *et al.*, 2022; **Marcec & Likic**, 2022; **Schalkwyk** *et al.*, 2020). Despite the risks on *Twitter*, **Haunschild and Bornmann** (2021) saw its potential as an early warning system for identifying potentially problematic information. Beyond *Twitter*, there are also other proposals. **Fraumann and Colavizza** (2022) reviewed and identified the important role that both news and blogs have played in science communication during the pandemic. In addition, **Colavizza** (2020) observed efforts by the *Wikipedia* community to incorporate the main research findings by referencing relevant publications.

The exceptionality of this situation resulting from the pandemic has thus been demonstrated, showing remarkable differences with other related phenomena, or already known patterns. There are differences between the various types of medical research outputs and the impact or attention they receive, such as the role played by preprints (**Majumder & Mandl**, 2020; **van Schalkwyk & Dudek**, 2022). Since the beginning of the pandemic, and despite the fact that it marked a period in which studies focused so much on a single topic, there was considerable concern that the content and quality of this research might not meet public health needs (**Odone** *et al.*, 2020). This concern eventually became a reality; it was found that the quality and evidence of the study types of many papers was below the usual standards (**Jung** *et al.*, 2021). Therefore, the COVID-19 publications could provide an opportunity to study whether the characteristics, special the type of medical research studies are related to the attention they receive in the main social media. In other words, the metric differences that may exist between, for example, a "Case Report" or a "Clinical Trial" could be studied.

This is possible because the field of Health Sciences has a classification of typologies, among which differences in scientific evidence and clinical value can be found (**Röhrig** *et al.*, 2009). Databases such as *Embase* or *Medline* classify their articles according to study type considering its design and using this information different studied have demonstrated that the study type is associated with the citation rates (**Okike** *et al.*, 2011; **Patsopoulos** *et al.*, 2005). For example, this phenomenon occurs with Systematic Reviews that received double the number of citations than Non-Systematic Reviews (**Bhandari** *et al.*, 2004; **Montori** *et al.*, 2003). Regarding altmetrics, a similar relationship has also been found between mentions and document type, as is the case with editorial materials, which have a high attention in social media despite bean rarely cited (**Haustein bi**, 2015). But there is no literature that explores the impact that the research study type and evidence level may have on altmetric mentions. Our main objective is to analyse whether the altmetric attention received by COVID-19 medical studies is associated with the research study type and evidence level. To achieve this main objective, the following specific objectives were set:

1. Calculate the most relevant altmetrics for the papers published on COVID-19 considering the type of study as the main variable.
2. Determine through different statistical tests if there are significant differences in the values observed in each type of study.
3. To perform a ranking of the different types of studies considering their altmetrics and compare them with the traditional evidence pyramids.

This paper is an considerable expansion of a preliminary study presented at the *STI 2022* (**Valderrama & Torres-Salinas**, 2022).

## 2. Methodology

We collected data from two sources: *PubMed* and *Altmetric.com*. Data were retrieved on 21 November 2022. Firstly, *PubMed* was used to retrieve the bibliographic records of the COVID-19 scholarly outputs published in the year 2021. Specifically, the search was carried out through *PubMed Clinical Queries* using the following query:

*(COVID-19[MeSH Terms] OR SARS-CoV-2[MeSH Terms] OR coronavirus [MeSH Terms]) AND ("2021/01/01"[Date - Publication]: "2021/12/31"[Date - Publication])*

This search resulted in a total of 93,024 publications that were classified according its study types that is assigned directly by the publishers or the Index Section in the NLM. For our aims the following study types were considered: 1. Case Reports; 2. Clinical Trials; 3. Consensus Development Conferences & Guidelines; 4. Reviews (all those that are Systematic Review are omitted); 5. Systematic Reviews; 6. Meta-Analyses; 7. Observational Studies. This reduced the total number of publications to 20,668. The distribution of publications by study type is unequal, with the majority being Reviews (9,873), followed in smaller numbers by Case Reports (4,254) and Observational Studies (3,117), after which are Systematic Reviews (2,101), Meta-Analyses (1,358) and Clinical Trials (1,325), and in last place Consensus Development Conferences & Guidelines (143).

Mentions were retrieved using *Altmetric.com* using the DOI. *Altmetric.com* had indexed 16,672 from *PubMed* having at least one mention. Regarding the selection of social media metrics it is necessary to point out a common problem in such studies—the unequal number of metrics counted by source (**Zahedi** *et al*., 2014). Frist, we removed some sources according to the following criteria: (a) Platforms with an irrelevant number of mentions (e.g., *YouTube* or *Stack Overflow*; (b) Platforms with a strong geographical component (e.g. *Weibo* and *Reddit*); and (c) Platforms that no longer exist or do not work (e.g. *LinkedIn*, *Google+*, *Sina Weibo* and *Pinterest*). Second, for ensure statically validity a threshold of 30% of publications with at least one mention was established. Finally, the metrics selected were: news mentions, *Twitter* mentions, and *Mendeley* readers. Likewise, *Dimension* citations were considered to undertake a comparison with traditional bibliometric indicators.

Descriptive statistics were used to study the distribution and stablish significant differences of altmetric mentions by type of study. First, we applied The Kruskal-Wallis test to contrast the hypothesis of equality of the medians between the variables to identify potential differences in performance (**Samuels** *et al*., 2011). Second, was to perform pair-by-pair comparisons by means of the Mann-Whitney and serves to check whether there are significant differences between two variables for this purpose, the test calculates the U. Finally, we used a test of independence based on the chi-square statistic on the contingency table of joint frequencies generated by Mann-Whitney test, so that when $p < 0.05$ we could conclude that there was a similarity between two variables, and therefore a similar level of evidence.

## 3. Results

### 3.1. Descriptive analysis

The cumulative sums of values by metric and study type are shown in **Figure 1,** where it can be clearly seen that Reviews have the highest values for all metrics except for news mentions. This can be explained by the fact that it is the most numerous study type, representing 49% of the publications with altmetrics that are studied. Following the Reviews, Clinical Trials stand out in *Altmetric Attention Score* (AAS) and *Twitter* mentions. In the rest of the study types, the altmetrics show a similar dynamic.

**Figure 1**. Cumulative sums of values by metric and study type of COVID-19 publications

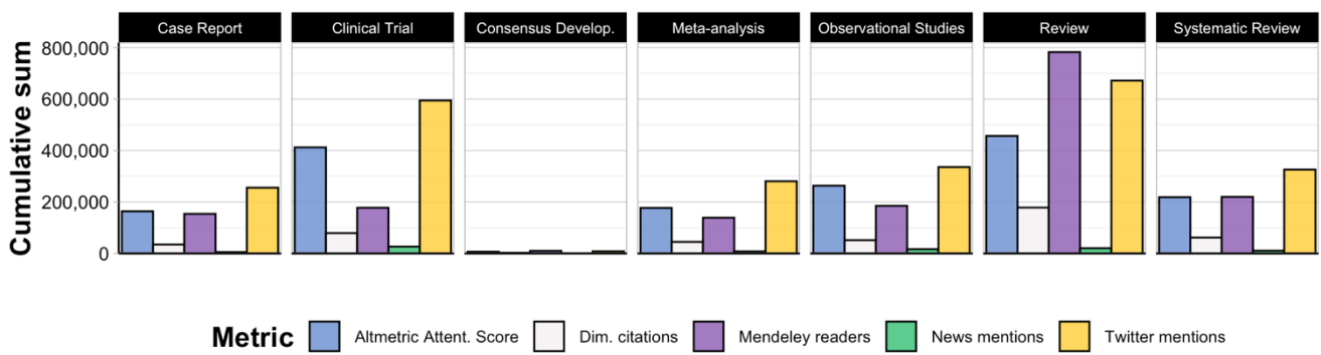

**Table 2**. Descriptive statistics of COVID-19 publications and their metrics by study type

| Study type according to type PubMed database | General Metrics | | Median values (interquartile range) | | | | |
|---|---|---|---|---|---|---|---|
| | Num. Publ. | Pub. with mentions | Altmetric Atten. Score | News mentions | Twitter mentions | Mendeley readers | Dim. citations |
| *1. Case Report* | 4,254 | 2,984 | 4 (1-14) | 0 (0-1) | 3 (1-16) | 37 (22-59) | 5 (1-12) |
| *2. Clinical Trial* | 1,325 | 1,169 | 12 (4-82) | 0 (0-3) | 10 (3-61) | 87 (51-152) | 13 (5-39) |
| *3. Meta-Analysis* | 1,358 | 1,196 | 10 (3-37) | 0 (0-1) | 10 (4-35) | 75 (43-137) | 17 (8-38) |
| *4. Observational Studies* | 3,117 | 2,460 | 7 (2-24) | 0 (0-1) | 6 (2-24) | 55 (32-90) | 9 (3-19) |
| *5. Consensus Develop.* | 143 | 117 | 14 (4-49) | 0 (0-2) | 15 (2-50) | 63 (35-113) | 9 (3-22) |
| *6. Systematic Review* | 2,101 | 1,885 | 9 (3-33) | 0 (0-1) | 10 (4-33) | 79 (46-136.75) | 15 (6-34) |
| *7. Review* | 9,873 | 8,177 | 7 (2-19) | 0 (0-1) | 5 (2-20) | 59 (34-104) | 9 (3-21) |
| *All* | 20,668 | 16,672 | 7 (2-22) | 0 (0-1) | 6 (2-24) | 57 (33-101) | 9 (3-22) |

In general, clinical trials and Consensus Development Conferences & Guidelines are the study types with the highest medians. It is noteworthy how Clinical Trials stand out in news mentions and *Mendeley* readers, while the Consensus Development Conferences & Guidelines have the highest median values in *Altmetric Attention Score* and *Twitter* mentions. However, it should be noted that the number of Clinical Trials (1325) is much higher than that of Consensus Development Conferences & Guidelines (143). Another study type that stands out in altmetrics are Meta-Analyses. With a median value of 10 in both *Altmetric Attention Score* and *Twitter* mentions, as well as the third highest median on *Mendeley* readers (75), this study type shows that its social attention is not restricted to a single altmetric source or specific social community. The altmetric values of Meta-Analyses follow very similar patterns to those of Systematic Reviews, the latter being the second type with the best median value in *Mendeley* readers (79). In contrast, there are the Case Reports, which, although they have many publications (4354), have the lowest medians for all altmetrics. Finally, it could be mentioned that Meta-Analyses and Systematic Reviews are not only the most cited study types but also have the highest medians in *Twitter* mentions and *Mendeley* readers.

### 3.2. Statistical differences between the types of studies

Comparisons are then made between the metrics of the seven study types to analyse how the selected metrics perform according to each study type. This comparison was done by means of the Kruskal-Wallis test, resulting in $p<0.001$; this means that there were significant differences between the metrics of each type of study. K-W test confirms the results observed in **Table 2** and **Figure 1** as indicates that altmetrics present very different values depending on the type of study. A cross-tabulation of coincidences between study type and metrics, collecting the p-values of the Mann-Whitney U test, is shown in **Table 3**. As can be seen, the altmetric indicator having the most coincidences within the study types is news mentions, whose number of coincidences is 7, followed by *Twitter* mentions with a total of 5 coincidences. *Mendeley* readers showed the lowest number of matches. Within the study types, the Consensus Development Conferences & Guidelines are found to match their p-values with at least 5 study types. It is followed by the Systematic Reviews and Observational Studies, each having coincidences with another 4 study types, including the grouping of both. It is noteworthy that the Systematic Review and Meta-Analysis groups have the same values for *Altmetric Attention Score*, *Twitter* mentions, news mentions and *Mendeley* readers. This indicate that these types of studies have a relevant role. Finally, the chi-square test of independence was applied. In this hypothesis test, the null hypothesis (H0) was that there is no relationship between study type and metrics, while the alternative hypothesis (H1) was that there is a relationship between study type and metrics. The test result was significant ($\chi^2$=294,569.85; p<0.001).

**Table 3**. Cross-tabulation of coincidences between altmetric indicators by study type, grouped two by two

| | Altmetric Attent. Score | Twitter mentions | News mentions | Mendeley readers | Dimensions citations |
|---|---|---|---|---|---|
| *Meta-analysis Systematic Review* | 0.415 | 0.831 | 0.932 | 0.192 | * |
| *Meta-analysis Consensus Develop.* | 0.229 | 0.377 | 0.325 | * | * |
| *Clinical Trial Consensus Develop.* | 0.415 | 0.990 | 0.090 | * | * |
| *Consensus Develop. Systematic Review* | 0.182 | 0.284 | 0.288 | * | * |
| *Observational Studies Consensus Develop.* | * | * | 0.086 | 0.084 | 0.814 |
| *Clinical Trial Systematic Review* | * | 0.094 | * | * | 0.137 |
| *Consensus Develop. Review* | * | * | * | 0.604 | 0.985 |
| *Meta-analysis Observational Studies* | * | * | 0.105 | * | * |
| *Observational Studies Systematic Review* | * | * | 0.072 | * | * |
| *Observational Studies Review* | * | * | * | * | 0.211 |

## 4. Discussion and conclusion

In this paper we focused on the altmetrics of COVID-19 studies published in 2021, taking the main types of medical studies to analyse the differences between them. Results indicate that altmetrics in Health Science research, specifically in the COVID-19 research front, could be highly determined by the type of research study, and alternatively suggest that altmetrics can capture the utility of the research explored here. In Medicine and especially in Evidence-Based Medicine, the usefulness of academic papers is linked to the evidence of their results and their practical application in the clinical world. One way to visualize utility is through the *Pyramid of Scientific Evidence* in which studies are assigned to levels of evidence based on their methodology, the evidence pyramid is an easy way to visualize this hierarchy of evidence the most valuable information (**Arsenault**, 2022). For example, in the **Figure 2A** a pyramid called by University of Washington Health Sciences Library (**Kowalczyk and Truluck;** 2013, **Murad** *et al.*, 2016) had been included. It can clearly be seen how the types of studies are ordered with the Consensus Development Conferences & Guidelines on the top. In this way (**Figure 2B**) a pyramid of evidence has been elaborated using the quantitative data obtained in results, specifically we have ordered study types using the values of the *Altmetric Attention Score* included in **Table 2**.

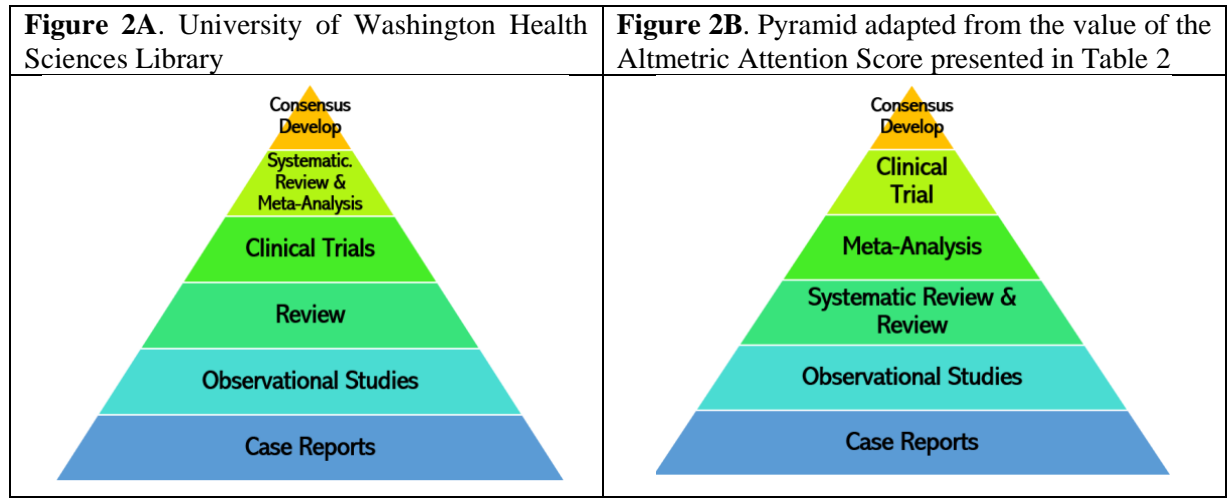

**Figure 2A**. University of Washington Health Sciences Library

**Figure 2B**. Pyramid adapted from the value of the Altmetric Attention Score presented in Table 2

As it can be seen both pyramids are essentially the same, with the main difference of the Clinical Trials, which in the case of Washington University are in third place and in the one offered by the *Altmetric Attention Score* in second place. If we compare the results generated quantitatively with altmetrics with other pyramids of evidence generated by specialists (**Arieta-Miranda** *et al.*, 2022; **Murad** *et al.*, 2016) the similarities are more than reasonable. This is explained by the fact that altmetric capture the social attention that publications receive, so the typologies closest to society, at least the most useful ones, are likely to receive more attention on social platforms. For example, Consensus Development Conferences & Guidelines are a way to bring together citizens, decision-makers and a series of experts to address issues of public importance, Clinical Trials are situated at the peak since their results are highly valid (**Lazcano-Ponce** *et al.*, 2004), Meta-Analysis is the statistical process of analysing and combining results from several similar studies (**Harris** *et al.*, 2014). Reviews (5), in addition to their educational component, are hypothesis generators, proving very important to analyse a new topic such as COVID-19 (**Valderrama** *et al.*, 2021).

We can conclude that, depending on the type of study, altmetrics reach different values and that, in addition, these values are able to capture the usefulness and evidence of those of the studies as we have seen when comparing our results with the pyramids of evidence. These are results that provide empirical evidence on the possible meaning of altmetrics and open the doors to their application in Evaluative Bibliometrics, at least in the field of Health Sciences. This study is not free of limitations. Altmetrics from only three social media were considered, one of them news mentions, which are present in approximately one third of the publications studied. Similarly, despite a high volume of COVID-19 publications, only a single year's publication period was used. For this reason, future work should explore this relationship between study type and altmetrics in the medical field beyond COVID-19 studies.

**Funding information**

This work was funded by the Spanish Ministry of Science and Innovation with grant number PID2019-109127RB-I00/SRA/10.13039/501100011033, and Regional Government of Andalusia Junta de Andalucía grant number A-SEJ-638-UGR20. Wenceslao Arroyo-Machado received an FPU Grant (FPU18/05835) from the Spanish Ministry of Universities.